\documentclass[aps,prl,twocolumn,superscriptaddress,showpacs,preprintnumbers,longbibliography]{revtex4-1}
\usepackage[colorlinks,bookmarks=false,citecolor=blue,linkcolor=red,urlcolor=blue]{hyperref}
\usepackage{physics}
\input pdfcolor.tex
\usepackage{colortbl,amsthm,txfonts}
\usepackage{verbatim}
\usepackage{graphicx}
\usepackage{epsfig}
\usepackage{xcolor}
\usepackage{dcolumn}
\usepackage{soul}
\usepackage{bm}

\usepackage{amsmath,amssymb}

\makeatletter
\renewcommand*\env@matrix[1][*\c@MaxMatrixCols c]{%

\hskip -\arraycolsep
\let\@ifnextchar\new@ifnextchar
\array{#1}}
\makeatother

\begin{document}

\title{Metal-insulator transition and thermal scales in $d$-wave altermagnet}
\author{Santhosh Kannan}
\thanks{These authors contributed equally to this work.}
\author{Jainam Savla}
\thanks{These authors contributed equally to this work.}
\author{Madhuparna Karmakar}
\email{madhuparna.k@gmail.com}
\affiliation{Department of Physics and Nanotechnology, 
SRM Institute of Science and Technology, 
Kattankulathur, Chennai 603203, India}

\date{\today}

\begin{abstract}
We present the first finite-temperature study of a strongly correlated $d$-wave altermagnet across the Mott insulator-metal transition using a non-perturbative numerical approach. We map out the thermal phase diagram and provide quantitative estimates of the transition scales in an interacting altermagnet. We show that altermagnetism-induced geometric frustration stabilizes a finite-temperature correlated magnetic metal and enhances the magnetic transition scale across regimes of interaction. These results establish the finite-temperature landscape of correlated altermagnets and clarify the role of strong electronic interactions in this phase.
\end{abstract}

\maketitle

\textit{Introduction:} Effectively considered to be the long sought after magnetic analogue of the unconventional superconductors, altermagnet (ALM) promises to capture the best of both ferromagnetism (FM) and antiferromagnetism (AFM) \cite{jungwirth_prx2022,pan_natrevmat2025,mason_annrev2025}. With their spin-split energy bands but zero net magnetization and spin-momentum locking sans heavy element dependent spin-orbit coupling (SOC), altermagnets are now looked forward to as the potential new candidate for spintronic devices and applications \cite{pan_natrevmat2025,mason_annrev2025}. The non-relativistic momentum ($k$) dependent spin splitting in the ALMs are dictated by the breaking of the $\mathcal{PT}$ symmetry, that lifts the Kramer's degeneracy. Following its initial proposal the recent upsurge of research on this new class of magnetism quickly confirmed the ALM order over a wide rage of metals, superconductors and insulators such as, RuO$_{2}$ 
\cite{sinova_sciadv2020,vasilyev_sciadv2024},FeSb$_{2}$ \cite{jungwirth_prx2022,smejkal_pnas2021}, CrSb \cite{jungwirth_prx2022,constantinou_natcom2024}, VNb$_{3}$S$_{6}$ \cite{jungwirth_prx2022}, CoNb$_{3}$S$_{6}$ \cite{sinova_sciadv2020}, La$_{2}$O$_{3}$Mn$_{2}$Se$_{2}$ \cite{ji_prm2026}, CoNb$_{4}$Se$_{8}$ \cite{ghimire_natcom2025}, KV$_{2}$Se$_{2}$O \cite{tian_prb2025} etc. Transport, spectroscopic and magnetic measurements carried out on these materials brought forth non trivial observations in the form of large giant magneto resistance (GMR) and tunneling magnetoresistance (TMR) \cite{xiao_prb2023}, anomalous Hall effect (AHE) \cite{zhu_natelec2022,mitchell_natcom2018,berger_prr2020} etc.,  to name a few.

Theoretical investigations on ALMs precede the experimental outcomes and are primarily categorized into: (i) first principal based band 
structure studies bringing forth the $k$-dependent spin-splitting and (ii) group theory based analysis of the crystalline symmetry, in candidate materials \cite{jungwirth_prx2022,pereira_prb2024,smejkal_pnas2021,brink_mattodayphys2023,lu_natscirev2025,smejkal_arxiv2023,sinova_prb2024,haule_prl2025,olsen_apl2024,mazin_scipost2024,yang_chemsci2024,yang_jamchemsoc2025,rhone_prm2025,stroppa_prl2025,zhou_prl2025,smejkal_arxiv2024,constantinou_natcom2024,brink_comphys2025,shen_prl2024,liu_natcom2025,ma_nanolet2025,qian_natphys2025,chen_natphys2025,mazin_prbl2023,kim_prl2024,sato_prb2024,jungwirth_nature2024,ji_prm2025,valenti_arxiv2025,sasaki_prr2025,cano_jap2025,seo_npjspintronics2025}. Attention to the interplay between strong electronic correlation and magnetic order in itinerant ALMs is more recent and has remained largely restricted to mean field based calculations on relevant microscopic Hamiltonians, baring few exceptions \cite{valenti_prb2024,scheurer_prr2024,capone_prbl2025,lu_arxiv2025,fernandes_arxiv2025,fernandes2_arxiv2025,franz_prl2025,thomale_prl2025}. Very recently density functional theory (DFT) and dynamical mean field theory (DMFT) when used in combination showed significant renormalization of the electronic band structure and spectroscopic signatures at the ground state of ALM materials, arising purely out of strong electronic correlations \cite{fernandes_arxiv2025}. A concrete understanding of the microscopic formalism of ALMs in terms of the competing 
and coexisting short range electronic correlations, their stability against fluctuations and their deviation beyond the standard notions of Fermi liquid (FL) description however, awaits.

\begin{figure}[t!]
\centering
\includegraphics[width=0.95\columnwidth]{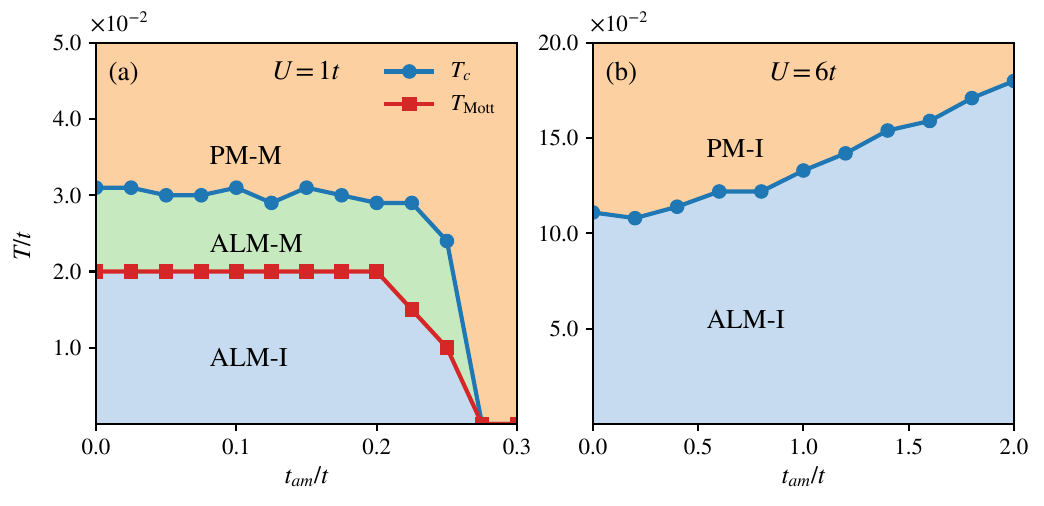}
\caption{Finite-temperature phase diagram in the $T-t_{am}$-plane at (a) weak $(U=t)$ and (b) strong $(U=6t)$ coupling. The thermal scales $T_{c}$ and $T_{Mott}$ quantify the loss of magnetic correlations and the collapse of the Mott gap, respectively.}
\label{fig1}
\end{figure}

In this letter, we analyze the physics of metal-insulator transition (MIT) in a ALM material based on a non-perturbative numerical approach and 
for the first time bring forth its thermal transition scales and phases. Our system is a two-dimensional (2D) lattice hosting $d_{x^{2}-y^{2}}$-wave ALM, while the fermionic interactions are modeled in terms of the prototypical Hubbard model. The choice of the numerical approach is the static path approximated (SPA) Monte Carlo technique which accounts for the spatial fluctuations and the corresponding short range correlations,  thereby provides accurate estimates of the thermodynamic phases and the transition scales therein. Quantified in terms of the thermodynamic and spectroscopic signatures our principal results encoded in Fig.\ref{fig1}, includes: (i) we provide the first accurate estimates of the thermal transition scales ($T_{c}$ and $T_{Mott}$) across MIT for a $d$-wave altermagnet in 2D, (ii) at weak electronic coupling, ALM stabilizes a finite temperature magnetically correlated metal, (iii) strong electronic coupling favors ALM correlations such that, $T_{c}$ increases monotonically with the ALM interaction strength. 

\textit{Theoretical formalism:} The Hubbard model on a square lattice with spin-dependent anisotropic hopping, reads as, 

\begin{eqnarray}
\hat H & = & \sum_{\langle ij\rangle, \sigma}(t_{ij} + \sigma t_{am}\eta_{ij} )(\hat c_{i, \sigma}^{\dagger} \hat c_{j,\sigma}+h.c)  \nonumber \\ && - \mu \sum_{i,\sigma}\hat n_{i,\sigma} + U \sum_{i}\hat n_{i, \uparrow}\hat n_{i,\downarrow}
\end{eqnarray}

\noindent where, $t_{ij}=t=1$ is the nearest neighbor hopping and sets the reference energy scale of the system. The 
second term depicts the $d_{x^{2}-y^{2}}$ ALM interaction such that, $t_{am}$ quantifies the strength of the interaction and 
$\eta_{ij}$ is the $d$-wave form factor, leading to $t_{\hat{x}}=-t - \frac{\sigma t_{am}}{4}$ and $t_{\hat{y}}=-t + \frac{\sigma t_{am}}{4}$; $\sigma = +(-)$ for the $\uparrow$($\downarrow$)-spin species. $U>0$ is the on-site Hubbard repulsion, $\mu$ is the chemical potential, adjusted to maintain a half-filled lattice. The model is made numerically tractable via Hubbard-Stratonovich (HS) decomposition of the interaction term, introducing the randomly fluctuating bosonic auxiliary fields ${\bf m}_{i}(\tau)$ and $\phi_{i}(\tau)$ which couples to the spin and the charge channels, respectively \cite{hs1,hs2} (see Supplementary Materials (SM) for the details). Our numerical approach is based on the adiabatic approximation wherein the slow, randomly fluctuating bosonic field serves as a static disordered background to the fast moving fermions, allowing for the bosonic fields to be treated as classical \cite{ciuchi_scipost2021,fratini_prb2023,kivelson_pans2023,karmakar_prml2025}. The approximation provides access to the real frequency ($\omega$) dependent quantities without requiring an analytic continuation.

Within the purview of SPA we retain the complete spatial fluctuations of ${\bf m}_{i}$ while $\phi_{i} = \langle n_{i}\rangle U/2$ is treated at the saddle point level (with $\langle n_{i}\rangle$ being the fermionic number density). The thermodynamic phases and phase transitions are quantified in terms of the: (i) static magnetic structure factor ($S({\bf q})$), (ii) single particle density of states (DOS) ($N(\omega)$), (iii) spin resolved electronic spectral function ($A_{\sigma}({\bf k}, \omega)$) and (iv) real space magnetic correlation (${\bf m}_{i}.{\bf m}_{j}$) (see SM). The results presented in this letter corresponds to a system size of $L=24 \times 24$ unless specified otherwise, and are found to be robust against finite system size effects (see SM).

\begin{figure}
\centering
\includegraphics[height=8.5cm,width=8.5cm,angle=0]{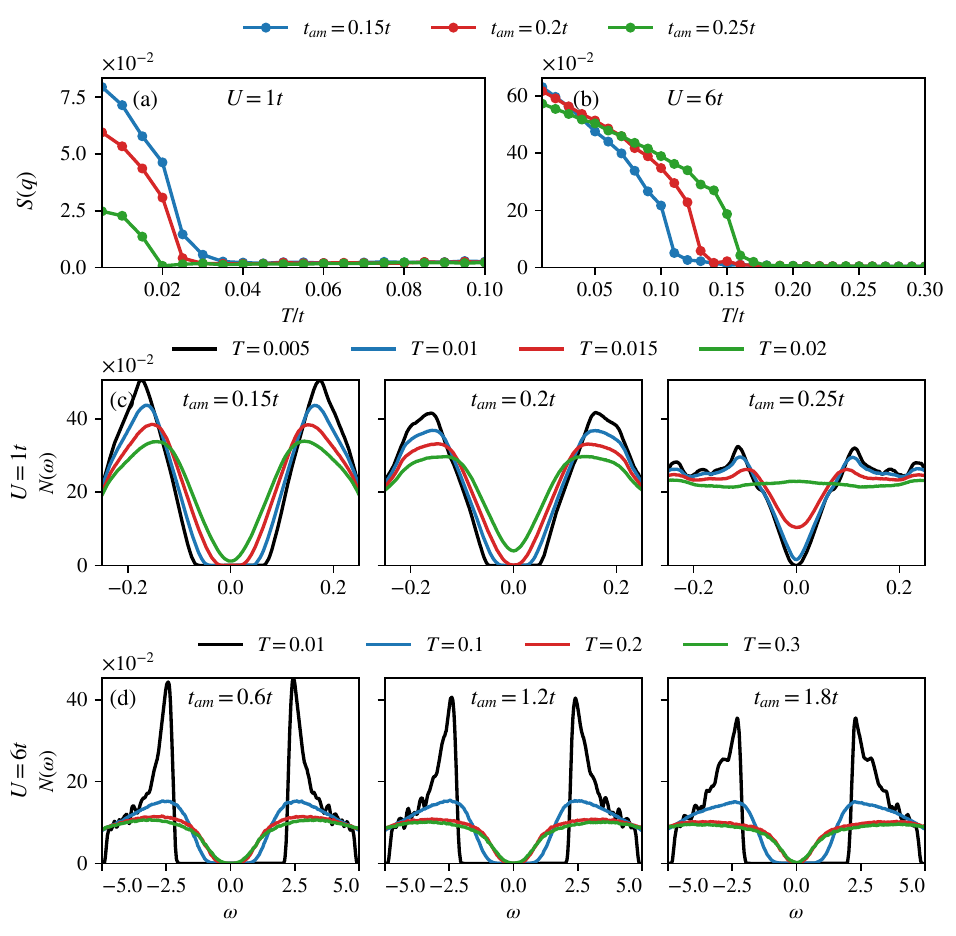}
\caption{Temperature dependence of the (a)-(b) static magnetic structure factor ($S(q=(\pi,\pi))$) and (c)-(d) single particle DOS ($N(\omega)$) 
at selected $t_{am}$ for $U=t$ and $U=6t$, respectively.}
 \label{fig2}
\end{figure}

\textit{Phase diagram and thermal scales:} Fig.\ref{fig1} constitutes the primary results of this work in terms of the ALM interaction-temperature ($t_{am}-T$) phase diagram in the (a) weak ($U=t$) and (b) strong ($U=6t$) coupling regimes, mapped out based on the observables shown in Fig.\ref{fig2}, viz. $S({\bf q})$, typifying the magnetic correlations and $N(\omega)$ with the corresponding spectral gap at the Fermi level being $E_{g}$. The thermodynamic phases are demarcated in terms of the thermal scales $T_{c}$ and $T_{Mott}$ quantifying the loss of the magnetic order and the MIT, respectively. The $T_{c}$ represents the Berezinskii-Kosterlitz-Thouless (BKT) transition scale with the corresponding magnetic state being (quasi) long range ordered in 2D \cite{mermin_wagner}. The thermodynamic phases are classified as: (a) altermagnetic Mott insulator (ALM-I) with $S({\bf q}) \neq 0$, $E_{g} \neq 0$, (b) altermagnetic metal (ALM-M) with $S({\bf q}) \neq 0$, $E_{g} = 0$, (c) paramagnetic metal (PM-M) with $S({\bf q}) = 0$, $E_{g} = 0$ and (d) paramagnetic insulator (PM-I) with $S({\bf q}) = 0$, $E_{g} \neq 0$. 
\begin{figure*}
\centering
 \includegraphics[height=14cm,width=15cm,angle=0]{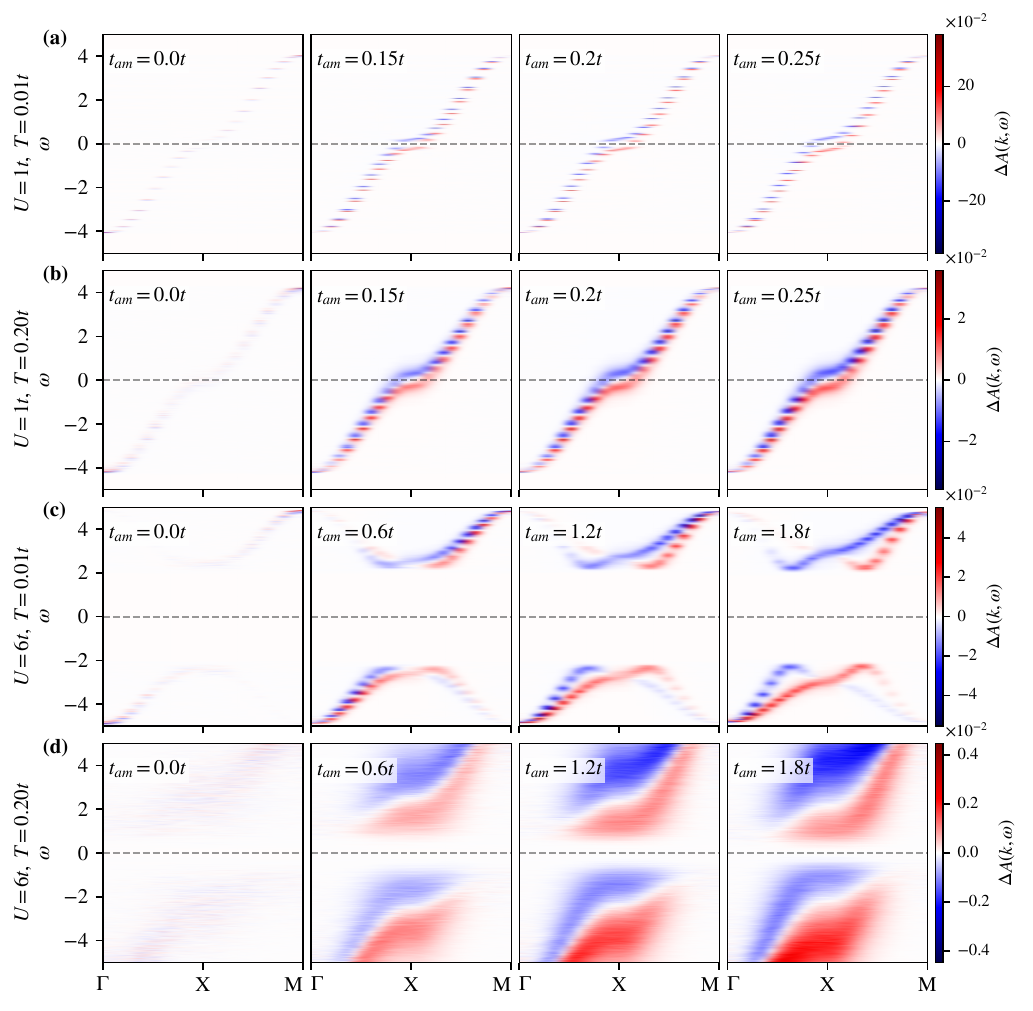}
 \caption{Spin-split spectral function $\Delta A(k, \omega) = A_{\uparrow} (k, \omega) - A_{\downarrow}(k, \omega)$ defined along the high symmetry trajectory $\Gamma - X - M$ for the representative weak $(U=t)$ and strong $(U=6t)$ coupling, at $T=0.01t$ and $T=0.20t$.}
 \label{fig3}
\end{figure*}

The magnetic correlation is collinear with $(\pi, \pi)$ Neel order irrespective of the choice of $U$ and $t_{am}$ (see SM for a comparison with the mean field state). Thermal fluctuations randomize the local moments leading to the progressive suppression of the (quasi) long range magnetic correlations, as shown in Fig.\ref{fig2}(a). The point of inflection of each $S({\bf q})$ curve quantifies the corresponding $T_{c}$ and the $T\gtrsim T_{c}$ regime typifies a magnetically disordered PM-M phase,  in the weak coupling regime. In terms of the spectroscopic signatures the ALM-I is quantified by a robust spectral gap at the Fermi level with prominent coherence peaks at the gap edges signifying the (quasi) long range magnetic correlations, as observed from Fig.\ref{fig2}(c), at $t_{am} = 0.15t$ and $t_{am} = 0.20t$. Thermal fluctuations lead to the progressive closure of the gap and broadens the coherence peaks via transfer of spectral weight away from the Fermi level, mapping out the MIT to ALM-M, across $T_{Mott}$.  The ALM-M regime at $T_{Mott} < T \lesssim T_{c}$ can be envisaged to be dominated by short range local magnetic correlations with isolated islands of $(\pi, \pi)$ order separated by magnetically disordered regions, evidenced by the gapless "V"-shaped $N(\omega)$ at the Fermi level. The limit of strong ALM interaction ($t_{am}=0.25t$) hosts a highly suppressed Mott gap at low temperatures which rapidly gives way to the ALM-M. The high temperature ($T\sim 0.20t$) phase at $t_{am}=0.25t$ is characterized by a featureless $N(\omega)$ attesting to the underlying state being a magnetically disordered PM-M.  At $t_{am} \neq 0 $ the PM-M phase hosts anisotropic spin-split electronic spectra 
conforming to the underlying $d$-wave ALM interaction, akin to the single particle spectra of the system at $U=0$ (see SM).   
\begin{figure}
\centering
\includegraphics[height=8cm,width=8cm,angle=0]{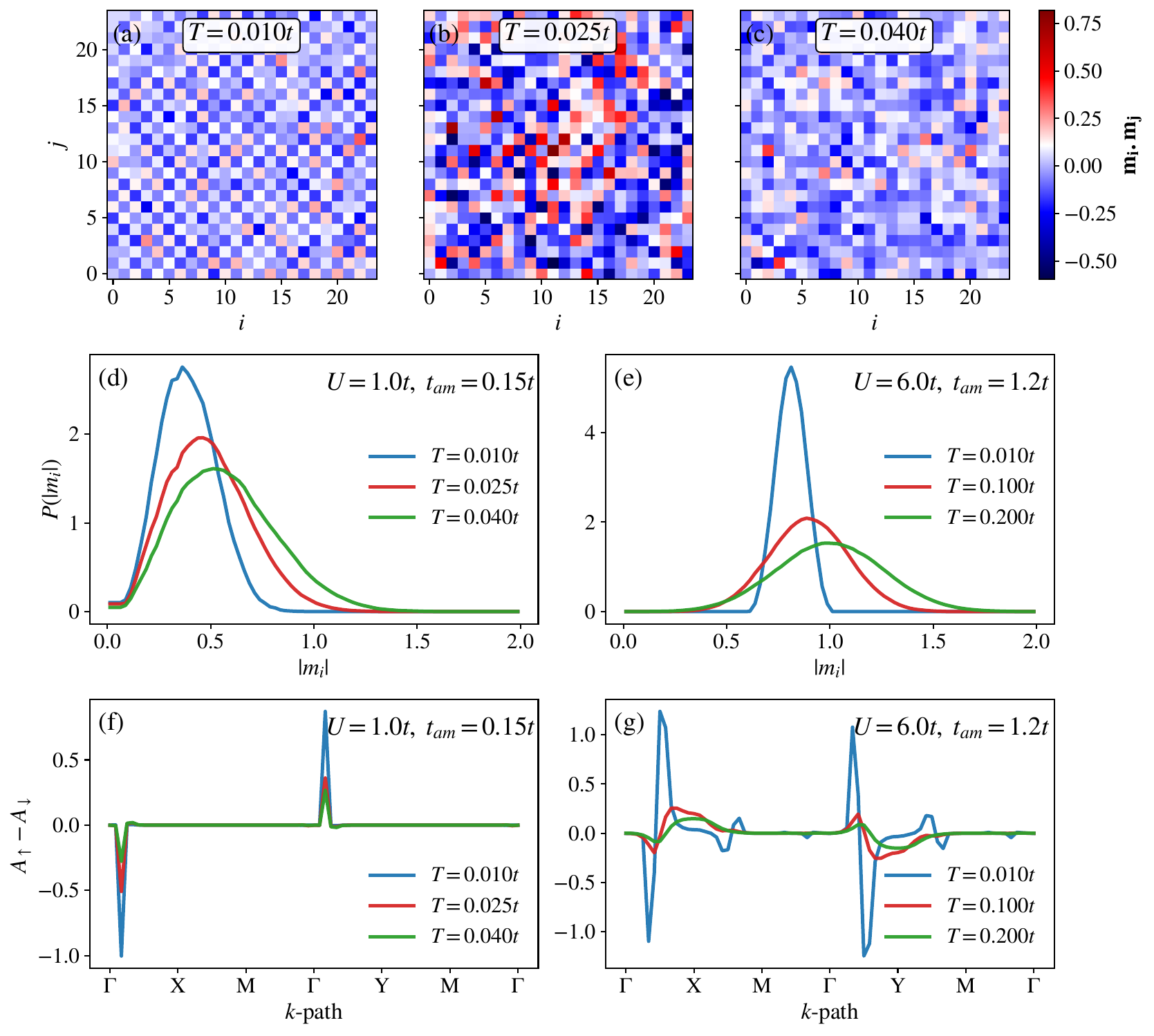}
\caption{(a)-(c) Real space maps (${\bf m}_{i}.{\bf m}_{j}$) representing the ALM-I, ALM-M and PM-M, respectively, at $U=t$ and $t_{am}=0.15t$, for a single Monte Carlo snapshot. Thermal evolution of the (d)-(e) local moment distribution and (f)-(g) ALM correlation ($\Delta A({\bf k})$) at $\omega \sim -3.5t$, for $U=t$ and $U=6t$, respectively, at selected $t_{am}-T$ cross sections.}
\label{fig4}
\end{figure}

The strong coupling regime ($U=6t$), as shown in Fig.\ref{fig1}(b), lacks any metallic phase and undergoes the transition between the 
ALM-I and PM-I across the $T_{c}$. Evidently, both these phases are quantified by a robust spectral gap ($E_{g} \propto U$) at the Fermi level 
and the large $U/t$ ensures that the ALM correlations survive over a significant range of $t_{am}$. In contrast to the weak coupling there is a monotonic {\it increase} in $T_{c}$ with $t_{am}$, opening up the possibility of stabilizing the ALM correlations in real materials by strong fermionic interactions. Fig.\ref{fig2}(d) shows that the single particle spectra at the low temperatures is characterized by van Hove singularities at the gap edges which progressively broadens via thermal fluctuations induced loss of (quasi) long range magnetic correlations.

For $T > T_{c}$, the $E_{g} \neq 0$ even though the (quasi) long range magnetic correlations are lost. We understand this as follows: The strong coupling regime is quantified in terms of the large local moments ${\bf m}_{i}$. For $T > T_{c}$ even though the angular coherence between the 
${\bf m}_{i}$'s is destroyed leading to the loss of the (quasi) long range magnetic order, the local moment amplitude $\vert m_{i}\vert$ doesn't get enough thermal window to undergo suppression. The PM-I regime therefore, corresponds to a region of randomly oriented large amplitude local moments, which opens up a finite spectral gap at the Fermi level, in a spirit similar to the bosonic insulators \cite{schilling_prb2018,kapitulnik_prb2008}.

\textit{Spectral function:} While $S({\bf q})$ and $N(\omega)$ are sufficient to quantify the magnetic correlations and Mott physics of the 
thermodynamic phases, the salient signature of ALM is the momentum dependent spin splitting of the electronic dispersion spectra. 
Fig.\ref{fig3} shows the thermal evolution of the spin resolved single particle spectral function splitting $\Delta A({\bf k}, \omega) = A_{\uparrow}({\bf k}, \omega)-A_{\downarrow}({\bf k}, \omega)$ along the high symmetry points at selected $U-t_{am}$ cross sections. By definition, $\Delta A({\bf k}, \omega) = 0$ at $t_{am} = 0$ irrespective of the choice of $U/t$. The topmost row represents the ALM-I at $U=t$ with a highly suppressed Mott gap and anisotropic spin-split electronic spectra, proportional to $t_{am}$. The Mott gap collapses at $t_{am} \sim 0.25t$, however, $\Delta A({\bf k}, \omega) \neq 0$ both at higher $t_{am}$ and at high temperatures, as observed from the second row of Fig.\ref{fig3}, typifying the PM-M phase.

Our observations on $\Delta A({\bf k}, \omega)$ at $U=6t$ are shown in the third and fourth rows of Fig.\ref{fig3}. At $T=0.01t$, representing the ALM-I, the system hosts a robust Mott gap at the Fermi level along with  $\Delta A({\bf k}, \omega) \neq 0$ exhibiting pronounced momentum dependent anisotropy. The high temperature phase ($T=0.20t$) is the representative of the PM-I, quantified by a weakly temperature dependent 
spectral gap at the Fermi level and thermally broadened, momentum dependent, spin-split Hubbard bands.

\textit{Discussion and conclusions:} This work maps out the thermodynamic phases and thermal scales of the prototypical $d_{x^{2}-y^{2}}$-wave ALM Mott insulator in 2D. Unlike the conventional Mott transition between an AFM Mott insulator and the magnetically disordered PM metal, thermal fluctuations stabilize a magnetically correlated ALM metal in the $d$-wave altermagnets, which we attribute to the {\it effective} geometric frustration arising out of the spin selective anisotropic hopping. The (quasi) long range ALM correlation is collinear, exhibiting a checkerboard pattern in the real space, as shown in Fig.\ref{fig4}(a) for ALM-I at $U=t$. Temperature randomizes the angular correlations between the local moments such that, the ALM-M phase comprises of isolated islands of short range correlations as shown in Fig.\ref{fig4}(b) and is characterized in terms of pseudogap-like signatures in the single particle spectra (see Fig.\ref{fig2}(b)). Our observations are in complete agreement with those of KV$_{2}$Se$_{2}$O \cite{tian_prb2025} and La$_{2}$O$_{3}$Mn$_{2}$Se$_{2}$ \cite{ji_prm2025}, the recently discovered $d$-wave altermagnets, analyzed based on optical and electron microscopy techniques, as well as DFT+DMFT calculations.  Note that ALM-M conforms to our understanding of the non-Fermi liquid metal in strongly correlated materials, characterized by a "V"-shaped single particle DOS; the  corresponding transport signatures are expected to deviate from the standard Fermi liquid description \cite{han_rmp2003,takagi_philmag2004}. The short range ALM correlations are eventually destroyed at higher temperatures and the resulting PM-M phase, as shown in Fig.\ref{fig4}(c) is essentially random.
 
Thermal fluctuations induced disordering of the local magnetic moments and the corresponding ALM correlations are shown in Fig.\ref{fig4}(d-g) 
at $U=t$ and $U=6t$,  at selected $t_{am}-T$ cross sections. Temperature suppresses the amplitude and broadens the distribution $P(\vert m_{i}\vert)$, shifting the mean magnitude of the local moment $\vert m_{i}\vert$ to larger values, indicating the loss of (quasi) long range order,  at $U=t$ (Fig.\ref{fig4}(d)). Stronger electronic interaction ($U=6t$) leads to a sharper distribution at low temperatures attesting the correlations to be longer ranged (Fig.\ref{fig4}(e)).

The ALM spin splitting quantified in terms of $\Delta A({\bf k}) = A_{\uparrow}({\bf k})-A_{\downarrow}({\bf k})$ at a selected energy of $\omega \sim -3.5t$ are shown in Fig.\ref{fig4}(f) and Fig.\ref{fig4}(g). Significant spin splitting along the $\Gamma-X$ and $\Gamma-Y$ paths, with the direction of polarization reversing between these paths (zero net magnetization (see SM)), confirms the $d_{x^{2}-y^{2}}$-symmetry of the ALM state. For $U=t$, the spin-splitting of the high temperature spectra continues to be ${\bf k}$-dependent even though $\Delta A({\bf k})$ is strongly suppressed (Fig.\ref{fig4}(f)) while at $U=6t$ the high temperature spectra is only weakly dependent on the momentum (Fig.\ref{fig4}(g)).

A variety of experimental probes have been implemented to characterize the ALM states. Spectroscopic techniques such as, soft X-ray angle resolved photo emission spectroscopy (SX-ARPES) \cite{jungwirth_nature2024,lee_prl2024,osumi_prb2024,hajlaoui_advmat2024,constantinou_natcom2024,lin_arxiv2024}, spin resolved ARPES \cite{liu_natcom2025,zeng_advsci2024,shen_prl2024,li_arxiv2024,lu_nanolett2025}, X-ray magnetic circular dichroism etc. were used to determine the spin dependent band splitting in $\alpha$-MnTe,  Rb and K doped V$_{2}$Se$_{2}$O$_{7}$ \cite{chen_natphys2025,qian_natphys2025,amin_nature2024,hariki_prl2024} etc.,  exhibiting clear $d$-wave symmetry. In a similar spirit, observations from spin dependent transport measurements on $\alpha$-MnTe \cite{chilcote_advfuncmat2024,gonzalez_prl2023,bey_arxiv2024}, thin films of Mn$_{5}$Si$_{3}$ \cite{reichlova_natcom2024,han_sciadv2024,kounta_prm2023,leiviska_prb2024,rial_prbl2024} and CrSb \cite{zhou_nature2024} etc. in terms of AHE \cite{chilcote_advfuncmat2024,gonzalez_prl2023,bey_arxiv2024,reichlova_natcom2024,han_sciadv2024,kounta_prm2023,leiviska_prb2024,rial_prbl2024,zhou_nature2024}, anomalous Nernst effect \cite{badura_natcom2025,chen_natphys2025}, magneto-optic Kerr effect (MOKE) \cite{dil_elecstruc2019,samanta_jap2020,zhou_prb2021,iguchi_prr2025,han_sciadv2024} etc. have put forward ALM as a potential candidate 
for spintronic applications. Recent reports exploring the impact of strong electronic correlations on the ALM phase have opened up the possibility of realizing this new class of magnetic ordering in a wide range of quantum materials, promising complex interplay of competing correlations and their functionalities \cite{tian_prb2025,ji_prm2025,fernandes2_arxiv2025}. 
 
The results presented in this letter are obtained based on the SPA Monte Carlo technique which retains the complete spatial fluctuations 
of the local moments and provides accurate estimates of the thermal transition scales and phases  \cite{mpk_imb,mpk_mass,dagotto_prl2005,rajarshi,nyayabanta_prb2016,nyayabanta_jpcm2017,nyayabanta_epl,mpk_spinliq,lieb_strain,karmakar_prml2025,shashi_kagome2024}. Suitably applied over a wide range of strongly correlated quantum systems such as, unconventional superconductors \cite{mpk_imb,mpk_mass,dagotto_prl2005}, Mott transition in geometrically frustrated materials \cite{rajarshi,nyayabanta_prb2016,nyayabanta_jpcm2017,nyayabanta_epl,mpk_spinliq}, quantum systems with flat electronic bands \cite{lieb_strain,karmakar_prml2025,shashi_kagome2024} etc. SPA works based on the adiabatic approximation of slow bosons. The primary artifact of the neglect of quantum fluctuations in this scheme is its restriction to access the temperature regime  $T < T_{FL}$, where $T_{FL}$ corresponds to the Fermi liquid coherence temperature, below which the translation invariance of the system is restored \cite{ciuchi_scipost2021,fratini_prb2023,kivelson_pans2023,karmakar_prml2025}.

\textit{Acknowledgment:} M.K. would like to acknowledge the use of the high performance computing facility (AQUA) at the Indian Institute of Technology, Madras, India.

\appendix
\section{Supplementary Information}

\setcounter{equation}{0}
\setcounter{figure}{0}
\renewcommand{\theequation}{S\arabic{equation}}
\renewcommand{\thefigure}{S\arabic{figure}}

\begin{figure*}[t]
\centering
\includegraphics[width=0.9\textwidth]{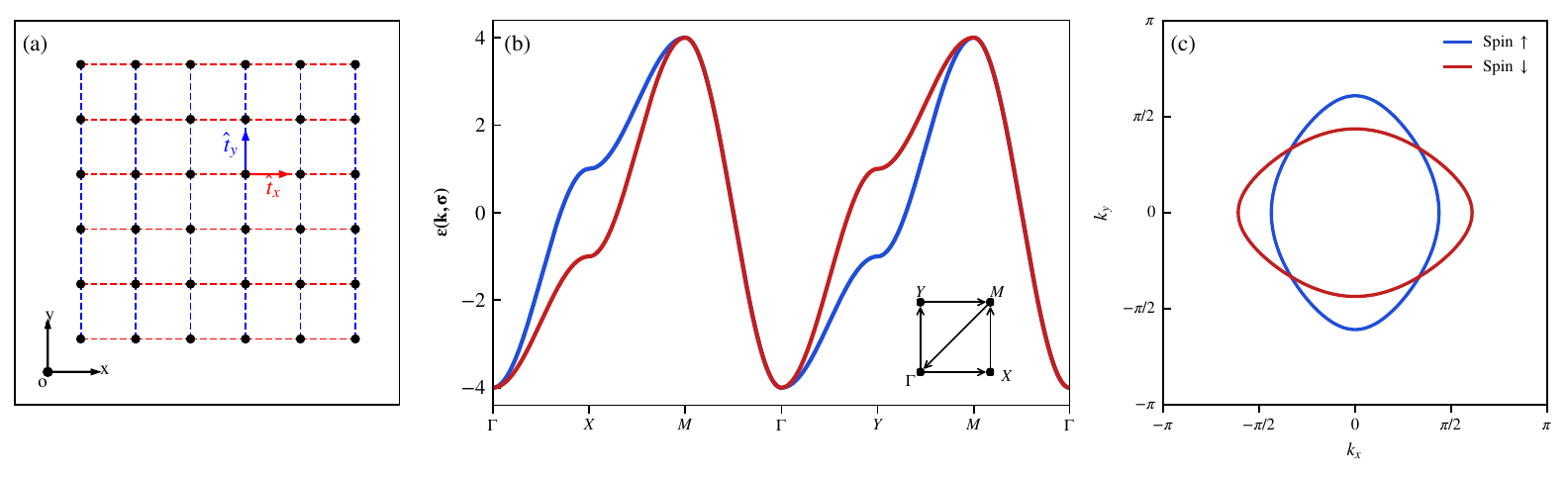}
\caption{
    (a) Schematic representation of the square lattice with $L = 6 \times 6$ with reference $x-y$ axes and the corresponding anisotropic hopping $\hat{t}_x$ and $\hat{t}_y$. (b) Tight binding energy dispersion of the model along high symmetry path $[\Gamma - X - M- \Gamma - Y - M - \Gamma]$ in reciprocal space with $U = 0t \:, t = 1.0 \:, t_{am} = 1.0t \:, \mu = -2.0$. (c) corresponding Fermi-Surface contours for spin $\uparrow$ and spin $\downarrow$ sectors.}
    \label{fig1_suppl}
\end{figure*}

\textit{Model and Formalism:} We model the system based on the prototypical Hubbard model on a 
square lattice with spin-dependent hopping mimicking the $d_{x^{2}-y^{2}}$-wave altermagnetic interaction, 
which reads as, 
\begin{eqnarray}
\hat H & = & \sum_{\langle ij\rangle, \sigma}(t_{ij} + \sigma t_{am}\eta_{ij} )(\hat c_{i, \sigma}^{\dagger} \hat c_{j,\sigma}+h.c)  \nonumber \\ && - \mu \sum_{i,\sigma}\hat n_{i,\sigma} + U \sum_{i}\hat n_{i, \uparrow}\hat n_{i,\downarrow}
\end{eqnarray}
\noindent where, $t_{ij}=t=1$ is the nearest neighbor hopping and sets the reference energy scale of the system. The 
second term depicts the $d_{x^{2}-y^{2}}$ ALM interaction such that, $t_{am}$ quantifies the strength of the interaction and 
$\eta_{ij}$ is the $d$-wave form factor, leading to $t_{\hat{x}}=-t - \frac{\sigma t_{am}}{4}$ and $t_{\hat{y}}=-t + \frac{\sigma t_{am}}{4}$; $\sigma = +(-)$ for the $\uparrow$($\downarrow$)-spin species, respectively (see Fig.\ref{fig1_suppl}(a)). The corresponding spin resolved single particle dispersion and the Fermi surfaces are shown in Fig. \ref{fig1_suppl}(b) and  Fig. \ref{fig1_suppl}(c), respectively. $U>0$ is the on-site Hubbard repulsion, while $\mu$ is the chemical potential, adjusted to maintain a half-filled lattice. 

We employ Hubbard-Stratonovich transformation \cite{hs1,hs2} to map our model onto an effective single-particle Mott-Hubbard spin-fermion Hamiltonian, enabling numerical calculations. Decoupling Hubbard term introduces the bosonic auxiliary fields viz. a vector field ${\bf m}_{i}(\tau)$ 
and a scalar field $\phi_{i}(\tau)$ that couples to the spin and charge densities,  respectively. The introduction of these bosonic auxiliary fields 
capture the Hartree-Fock theory at the saddle point, preserves the spin rotation symmetry and the Goldstone modes. In terms of the Grassmann 
fields $\psi_{i\sigma}(\tau)$, we write,
\begin{eqnarray}
\exp[U\sum_{i}\bar\psi_{i\uparrow}\psi_{i\uparrow}\bar\psi_{i\downarrow}\psi_{i\downarrow}] & = & \int \prod_{i}
\frac{d\phi_{i}d{\bf m}_{i}}{4\pi^{2}U}{\exp}[\frac{\phi_{i}^{2}}{U}+i\phi_{i}\rho_{i}+ \nonumber \\ && \frac{m_{i}^{2}}{U} -2{\bf m}_{i}.{\bf s}_{i}]
\end{eqnarray}

where, the charge and the spin densities are defined as, $\rho_{i} = \sum_{\sigma}\bar\psi_{i\sigma}\psi_{i\sigma}$ 
and ${\bf s}_{i}=(1/2)\sum_{a,b}\bar \psi_{ia}{\bf \sigma}_{ab}\psi_{ib}$, respectively. The corresponding partition function takes the form,
\begin{equation}
\begin{aligned}
\mathcal{Z} &= \int \prod_{i} \frac{d\bar{\psi}_{i \sigma} d\psi_{i \sigma} d\phi_{i} d\textbf{m}_{i}}{4\pi^2 U} \exp\left[ -\int_{0}^{\beta} \mathcal{L}(\tau) \, d\tau \right]
\end{aligned}
\end{equation}

with the Largrangian $\mathcal{L}$ being defined as, 

\begin{eqnarray}
{\cal L}(\tau) & = & \sum_{i\sigma}\bar\psi_{i\sigma}(\tau)\partial_{\tau}\psi_{i\sigma}(\tau) + H_{0}(\tau) 
\nonumber \\ && +\sum_{i}[\frac{\phi_{i}(\tau)^{2}}{U}+(i\phi_{i}(\tau)-\mu)\rho_{i}(\tau)+
\frac{m_{i}(\tau)^{2}}{U} \nonumber \\ && -2{\bf m}_{i}(\tau).{\bf s}_{i}(\tau)]
\end{eqnarray}

Here,  $H_0(\tau)$ is single particle kinetic energy contribution. Decoupling the quartic $\psi$ integral to quadratic costs an additional integration over the bosonic fields $\textbf{m}_i(\tau)$ and $\phi(\tau)$. The weight factor for these fields can be determined by integrating out $\psi$ and $\bar{\psi}$ fields. Using these weighted configurations one can determine the fermionic properties. Formally, 
\begin{equation}
\begin{aligned}
{\cal Z} = \int {\cal D}{\bf m}{\cal D}{\phi}e^{-S_{eff}\{{\bf m},\phi\}}    
\end{aligned}
\end{equation}

\begin{equation}
\begin{aligned}
S_{\mathrm{eff}}\{\mathbf{m},\phi\}
=& \ln \big[\mathcal{G}^{-1}\{\mathbf{m},\phi\}\big]
+ \sum_{i}\int_{0}^{\beta} d\tau
\left(
\frac{\phi_{i}^{2}}{U}
+ \frac{m_{i}^{2}}{U}
\right)
\end{aligned}
\end{equation}
where, ${\cal G}$ is the fermionic Green's function in a $\{{\bf m}_{i},\phi_{i}\}$ fluctuating background, given by,

\begin{eqnarray}
[{\mathcal G}^{-1}\{{\bf m}, \phi\}]_{(i\sigma, j\sigma^{\prime})} & = & [(\partial_{\tau}-\mu + i\phi_{i}(\tau))\delta_{ij}\delta_{\sigma \sigma^{\prime}} - (t_{ij} + \nonumber \\ && \sigma t_{am}\eta_{ij})\delta_{\sigma\sigma^{\prime}} - \frac{U}{2}{\bf m}_{i}(\tau).{\bf s}_{\sigma\sigma^{\prime}}\delta_{ij}] \nonumber \\ && \times \delta(\tau-\tau^{\prime})
\end{eqnarray}

The weight factor for an arbitrary space-time configuration $\{{\bf m}_{i}(\tau), \phi_{i}(\tau)\}$ involves computing the fermionic determinant in that background. If we express the auxiliary fields in terms of their Matsubara modes ${\bf m}_{i}(i\Omega_{n})$ and $\phi_{i}(i\Omega_{n})$, the various approximations can be recognized and compared \cite{karmakar_prml2025}.
\begin{figure*}
    \centering
    \includegraphics[width=\textwidth]{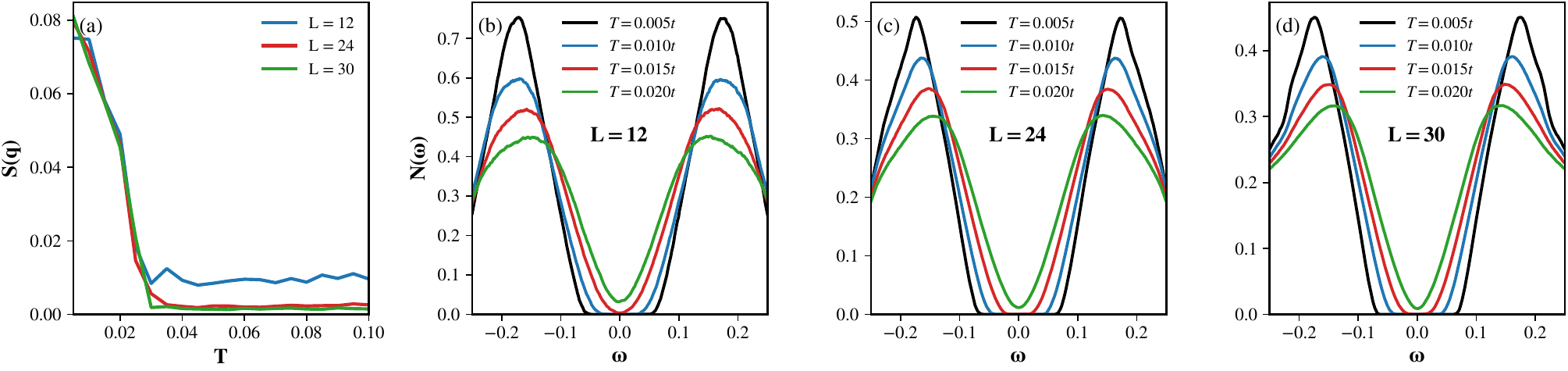}
    \caption{Finite system size effects at $U = t$ and $t_{am} = 0.15t$. Panels represents the thermal evolution of (a) $S({\bf q})$ at selected $L = 12, 24, 30$ and (b)-(d) $N(\omega)$ at selected values of $T = [0.005t, 0.010t, 0.015t, 0.020t]$ for each $L$, respectively. }
    \label{fig2_suppl}
\end{figure*}

As the tool of our choice we use static path approximated (SPA) Monte Carlo approach based on the adiabatic approximation, wherein the fast moving fermions are subjected to the random fluctuating background of the slow bosons. This allows us to treat the auxiliary fields as classical, and provides access to the real frequency dependent quantities without analytic continuation. In terms of Matsubara frequency this is tantamount to retaining $(\Omega_{n}=0)$ zero mode. Further, within the purview of SPA we treat $\phi_{i}$ at the level of its saddle point $\phi_{i}(\tau)=\frac{\langle n_{i}\rangle U}{2}$ where $\langle n_{i}\rangle$ is the fermionic number density. The effective Hamiltonian is therefore 
a spin-fermion model  wherein the fermions are coupled to the spatially fluctuating random background of bosonic classical field $\textbf{m}_{i}$ and reads as,

\begin{equation}
\begin{split}
H_{\mathrm{eff}}
=& -\sum_{\langle ij\rangle,\sigma}
\Bigl(t_{ij}+\sigma t_{am}\eta_{ij}\Bigr)
\left(\hat c_{i\sigma}^{\dagger}\hat c_{j\sigma}
+ \mathrm{h.c.}\right) \\
&- \tilde{\mu}\sum_{i\sigma}\hat n_{i\sigma} - \frac{U}{2}\sum_{i}\mathbf{m}_{i}\!\cdot\!\mathbf{s}_{i}
+ \frac{U}{4}\sum_{i} m_{i}^{2} .
\end{split}
\label{eq:H_eff}
\end{equation}
where, $\tilde \mu =\sum_{i}(\mu-\langle n_{i}\rangle U/2)$, and the last term of $H_{eff}$ corresponds to the 
stiffness cost associated with the classical field 
${\bf m}_{i}$. Here, ${\bf s}_{i}=\sum_{a,b}c_{ia}^{\dagger}{\bf \sigma}_{ab}c_{ib}$.
 
The random background configurations of $\{{\bf m}_{i}\}$ are generated numerically via Monte Carlo simulations and follow the Boltzmann distribution,
\begin{equation}
\begin{aligned}
P\{{\bf m}_{i}\} \propto Tr_{c,c^{\dagger}}e^{-\beta H_{eff}}
\end{aligned}
\end{equation}
\begin{figure}
\centering
\includegraphics[width=\columnwidth]{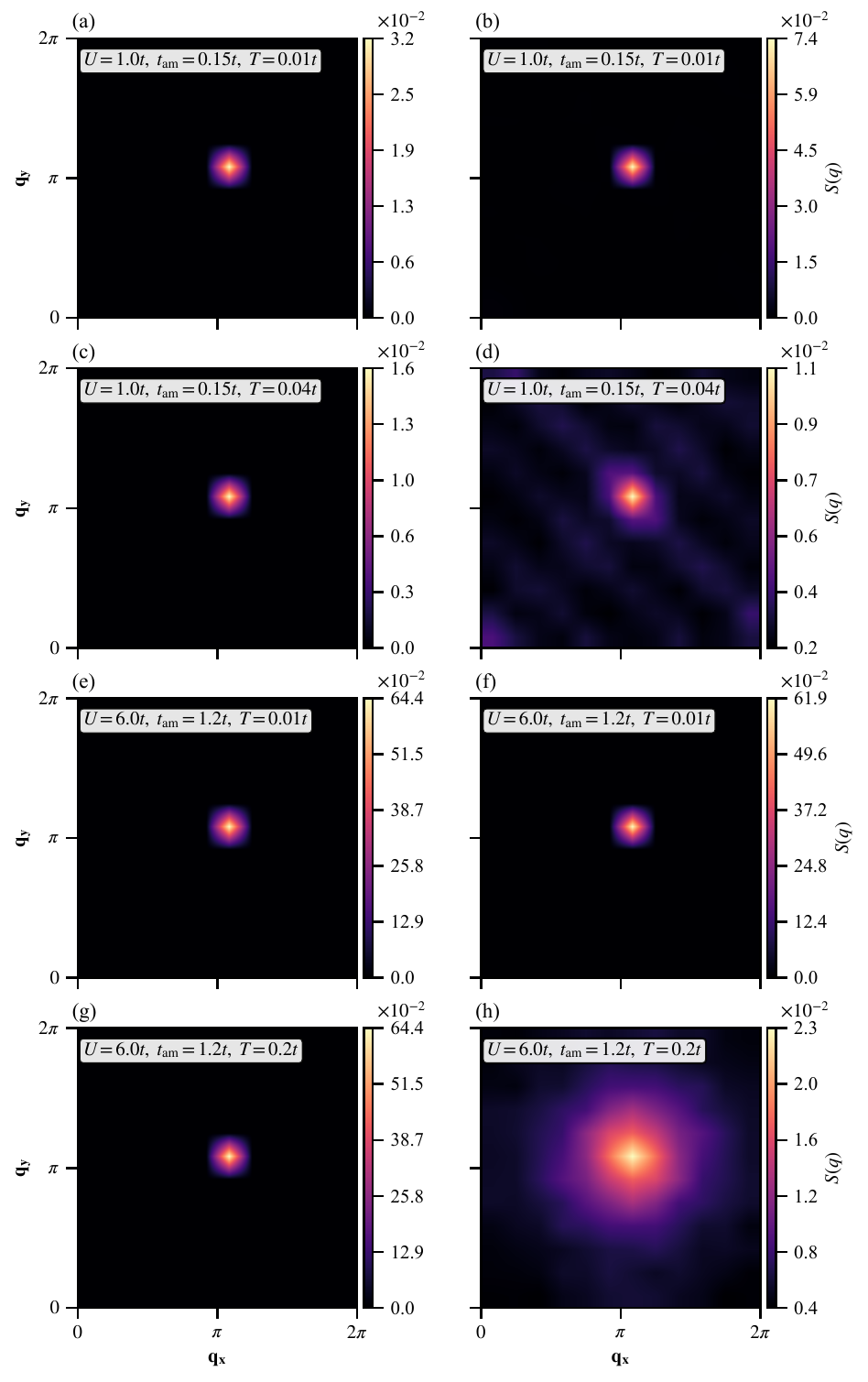}
\caption{Comparison of static magnetic structure-factor maps $S({\bf q})$: (a),(c) MFT; (b),(d) SPA at $U = t, t_{am} = 0.15t$ and 
$T=0.01t$, $T=0.04t$, respectively. (e),(g) MFT; (f),(h) SPA at $U = 6t, t_{am} = 1.2t$ and $T=0.01t$, $T=0.2t$, respectively.}
\label{fig3_suppl}
\end{figure}

\textit{Mean field theory:}
$H_{eff}$ is treated within the Hartree--Fock mean-field approximation by restricting the auxiliary fields to their saddle-point configurations. 
The saddle-point solutions are obtained by extremizing the effective action $S_{\mathrm{eff}}[\phi,\mathbf m]$ w. r. t these 
static fields, yielding the mean-field self-consistency conditions
\begin{equation}\label{eq:saddle_point}
\left.\frac{\delta S_{\mathrm{eff}}}{\delta\phi_i}\right|_{\substack{\phi_i(\tau)=\phi_i^{\mathrm{sp}}\\\mathbf m_i(\tau)=\mathbf m_i^{\mathrm{sp}}}}
\;=\;
\left.\frac{\delta S_{\mathrm{eff}}}{\delta\mathbf m_i}\right|_{\substack{\phi_i(\tau)=\phi_i^{\mathrm{sp}}\\\mathbf m_i(\tau)=\mathbf m_i^{\mathrm{sp}}}}
\;=\;0,
\end{equation}
which determine the Hartree (charge) shift $\phi_i^{\mathrm{sp}}$ and the local spin expectation value $\mathbf m_i^{\mathrm{sp}}$. Solving Eq.~\eqref{eq:saddle_point} explicitly leads to
\[
\phi_i = \frac{U}{2} \sum_\sigma \big\langle  \hat{c}^\dagger_{i\sigma} \hat{c}_{i\sigma} \big\rangle_{H_{MF}}, 
\quad 
\mathbf{m}_i = \frac{1}{2} \sum_{\alpha\beta} \big\langle \hat{c}^\dagger_{i\alpha} \boldsymbol{\sigma}_{\alpha\beta} \hat{c}_{i\beta} \big\rangle_{H_{MF}}.
\]
The self consistency of $\phi_{i}$ ensures a half filled lattice.

\textit{Indicators:} The different phases are classified and probed based on the following fermionic correlators computed on the 
equilibrium configuration of ${\bf m}_{i}$,

\begin{enumerate}
    \item Magnetic Structure factor:
    \begin{align}
    S(\mathbf{q}) \;=\; \frac{1}{N^2} \sum_{i,j} \langle \mathbf{m}_i . \mathbf{m}_j\rangle \, e^{i \mathbf{q} \cdot (\mathbf{r}_i - \mathbf{r}_j)} \label{eq:magnetic-structure-factor}
\end{align}
where $\mathbf{q}$ is the magnetic ordering wave vector and $N$ is the total number of lattice sites, $\langle \mathbf{\cdot} \rangle$ corresponds to MC configurational average.

\item Single Particle DOS:
\begin{align}
    N(\omega) \;=\;\frac{1}{N}\sum_{n, \sigma}\langle \delta(\omega-\epsilon_{n})\rangle
\end{align}    
where, $\epsilon_{n}$ correspond to the eigenvalues of the single equilibrium configuration.

\item Spin resolved spectral function:
\begin{align}
A_{\sigma}(\mathbf{k},\omega) &= -\frac{1}{\pi}\,\operatorname{Im} G_{\sigma}(\mathbf{k},\omega) \label{eq:spectral} \\
G_{\sigma}(\mathbf{k},\omega) &= \lim_{\delta \to 0^+} G_{\sigma}(\mathbf{k},i\omega_n)\big|_{i\omega_n \to \omega + i\delta} \notag
\end{align}
Here, \(G_{\sigma}(\mathbf{k},i\omega_n)\) is the imaginary-frequency Green's function, i.e. the Fourier transform of \(\langle \mathcal{T}_{\tau}\, c_{\mathbf{k}\sigma}(\tau)c_{\mathbf{k}\sigma}^{\dagger}(0)\rangle\) in imaginary time.
\end{enumerate}	
\begin{figure}
    \centering
    \includegraphics[width=\columnwidth]{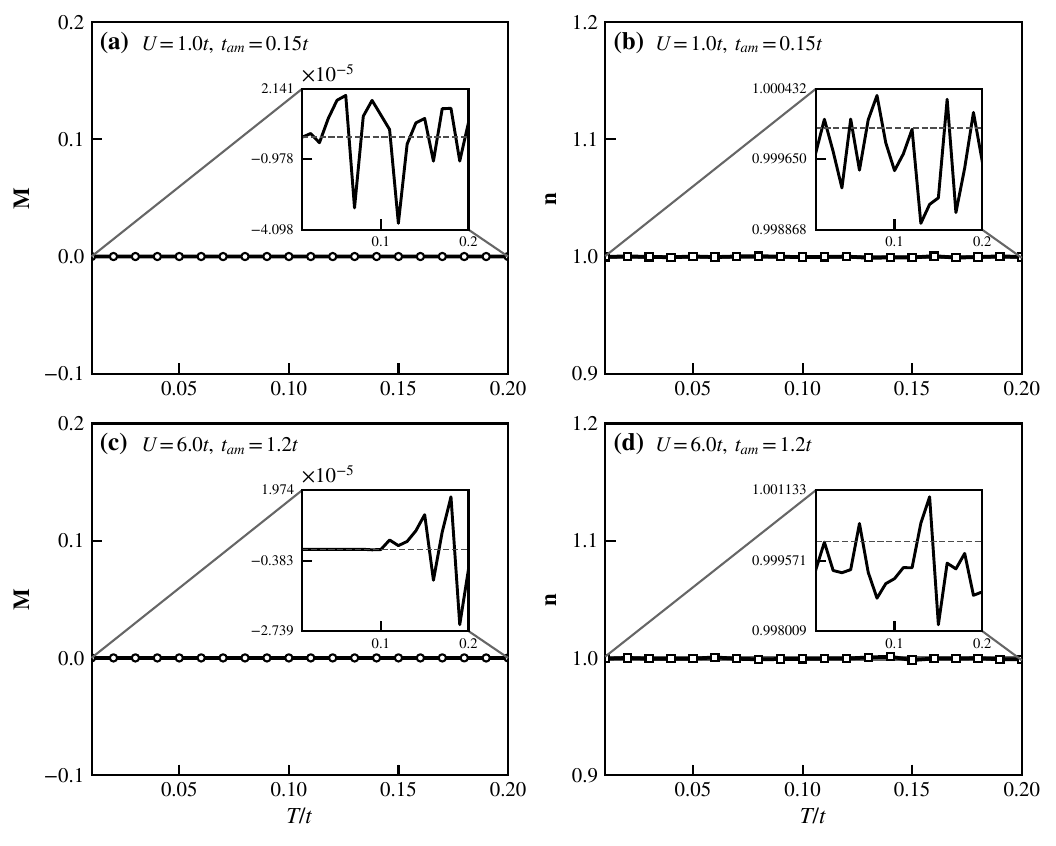}
    \caption{Temperature evolution of (a), (c) Net Magnetization $\langle n_\uparrow - n_\downarrow \rangle$ and (b), (d) total particle filling $\langle n_\uparrow + n_\downarrow \rangle$ at selected values of $U - t_{am}$ cross section. The inset in every panel shows a close look of the thermal fluctuations.}
    \label{fig4_suppl}
\end{figure}

\textit{Finite system size effect:} Fig.\ref{fig2_suppl} summarize the finite system size effects on the results discussed in this manuscript.  We compare the thermal evolution of $S({\bf q})$ and $N(\omega)$ for the selected values at $U = t$ and $t_{am} = 0.15t$ across different system sizes of $L = 12, 24, 30$. Fig. \ref{fig2_suppl}(a) shows that $S({\bf q})$ is largely immune to the system size except for $L=12$ and the corresponding $T_{c}$ scale is robust against the finite system size effect.  The corresponding single particle DOS at different system sizes are presented in Fig.\ref{fig2_suppl}(b-d) and is found to be largely immune to the system size effect, both in terms of the size of the spectral gap at the Fermi level and the thermal scale $T_{Mott}$ quantifying the MIT in this system.

\textit{Structure Factor Benchmark:} SPA maps on to the mean field theory (MFT) as the thermal fluctuations die down. In Fig.\ref{fig3_suppl} we present the comparison between the results obtained via MFT and SPA at selected $U-T-t_{am}$ cross sections. Fig.\ref{fig3_suppl}(a) and (c) shows the $S({\bf q})$ results as obtained via MFT at $U=t$, $t_{am}=0.15t$ for $T=0.01t (T<T_{c})$ and $T=0.04t (T > T_{c})$, respectively,  while for the same set of parameters the corresponding SPA results are presented in Fig.\ref{fig3_suppl}(b) and (d). We note that at these $U-t_{am}$ cross sections the results obtained via SPA are in complete agreement with those of MFT, both is terms of the magnetic ordering wave vector as well as the correlation amplitude $S({\bf q})$.

The results can be compared with those shown in Fig.\ref{fig3_suppl}(e)-(h), representing the strong coupling regime at $U=6t, t_{am}=1.2t$. While the $T<T_{c}$ results at $T=0.01t$ (Fig.\ref{fig3_suppl}(e)-(f)) exhibit perfect agreement between the outcomes of MFT (Fig.\ref{fig3_suppl}(e)) and SPA (Fig.\ref{fig3_suppl}(f)), the comparison at $T=0.20t (T> T_{c})$ (Fig.\ref{fig3_suppl}(g)-(h)) shows gross over estimation of the $T_{c}$. The results obtained via SPA shows that thermal fluctuations rapidly suppress the magnetic correlations leading to a highly broadened and diffused $S({\bf q})$ peak as compared to the results obtained using MFT which exhibits a robust $S({\bf q})$ even at high temperature. Our observations confirm that the MFT can suitably capture the weak coupling regime even at high temperatures where the suppression of the order parameter amplitude dictates the loss of order. Away from the weak coupling regime the MFT fails to account for the loss of angular coherence between the local moments via thermal fluctuations that leads to the collapse of the magnetic order.

\textit{Magnetization:} In Fig.\ref{fig4_suppl} we present additional evidence supporting the realization of altermagnetism in our system by analyzing the temperature evolution of spin-resolved number densities $(n_{\uparrow} , n_{\downarrow})$, from which the net magnetization 
$M = n_{\uparrow} - n_{\downarrow} $ and total particle density $n = n_{\uparrow} + n_{\downarrow}$ were obtained. For all the specified values 
of $U/t$ and $t_{am}/t$, the system exhibits a vanishing net magnetization $(M = 0)$ throughout the temperature range which is a hall mark of ALM order. As we make a closer look into the inset in panel (a) and (c) which reveals that the thermal fluctuation in magnetization occurs only for $T > T_c$ until which $M$ stays exactly zero suggesting the robustness of the ordered phase. Simultaneously $n = 1$ across both weak and strong coupling indicates that the system is maintained at the half-filling.

\bibliography{mottalt.bib}
\end{document}